\documentstyle[prl,aps,multicol,epsf]{revtex}
\begin{document}
\draft
\title{Electronic Entanglement in the Vicinity of a
Superconductor}
\author{Gordey B.\ Lesovik$^{a,b,c}$, Thierry Martin$^b$, and
Gianni   Blatter$^{c}$}

\address{$^{a}$L.D.\ Landau Institute for Theoretical Physics, Russian
  Academy of Sciences, Kosygina Str.\ 2, 117940, Moscow}

\address{$^b$Centre de Physique Th\'eorique et Universit\'e de la
M\'editerran\'ee, Case 907, 13288 Marseille, France}

\address{$^c$Theoretische Physik, ETH-H\"onggerberg, CH-8093 Z\"urich,
  Switzerland}

\date{\today}
\maketitle

\begin{abstract}
  {A weakly biased normal-metal--superconductor junction is considered
    as a potential device injecting entangled pairs of quasi-particles
    into a normal-metal lead. The two-particle states arise from
    Cooper pairs decaying into the normal lead and are characterized
    by entangled spin- and orbital degrees of freedom. The separation
    of the entangled quasi-particles is achieved with a fork geometry
    and normal leads containing spin- or energy-selective filters.
    Measuring the current--current cross-correlator between the two
    normal leads allows to probe the efficiency of the entanglement
    (cond-mat/0009193).}
\end{abstract}
\maketitle

\begin{multicols}{2}
%\narrowtext

The nonlocal nature of
quantum mechanics has been demonstrated theoretically \cite{EPR}
using entangled pairs of particles several decades ago. Recently,
potential applications of this entanglement have been found in
quantum cryptography\cite{crypt}, in quantum
teleportation\cite{telep}, and in quantum computing\cite{Steane}.
It is thus necessary to search for practical ways to produce such
pairs given a specific interaction between particles. While past
experiments have focused~on~pairs of photons\cite{ent_photons}
propagating in vacuum, attention is now turning to electronic
systems \cite{Burkard_Loss}, where this entanglement interaction
can be stronger while coherence can still be maintained over
appreciable distances in mesoscopic conductors. A scheme was
recently presented\cite{Loss_Sukhorukov} which discussed the
entanglement of electrons via the exchange interaction in pairs of
quantum dots. Here, we propose a rather robust electronic
entanglement scheme based on the Andreev reflection of electrons
and holes at the boundary between a normal metal and a
superconductor.

The basic concept underlying the microscopic description of
superconductivity is the formation of Cooper pairs. A normal metal in
vicinity to a superconductor bears the trace of this phenomenon
through the presence of Bogoliubov quasi-particles, or through the
non-vanishing of the Gor'kov Green function \cite{Gorkov} $F = \langle
c_{{\bf k} \uparrow} c_{-{\bf k} \downarrow}\rangle$ ($c_{{\bf k}
  \sigma}$ denote the usual electron annihilation operators).  While
in a superconductor $F = \Delta/\lambda$ is a consequence of a nonzero
gap parameter $\Delta$ ($\lambda$ is the pairing potential), the
coherence surviving in the adjacent normal metal can be understood
through the presence of evanescent Cooper pairs.  These involve two
electrons with entangled spin- and orbital degrees of freedom,
carrying opposite spins in the case of usual $s$-wave pairing and with
kinetic energies above and below the superconductor chemical
potential. This proximity effect has been illustrated in several
recent experiments\cite{proximity}.

In order to detect this entanglement and implement it for
applications, it is necessary to achieve a spatial separation between
the two constituent electrons. The entanglement apparatus which is
proposed here consists of a mesoscopic normal-metal--superconductor
(NS) junction with normal leads arranged in a fork geometry (see Fig.\
1). Using appropriate spin- or energy-selective filters in the two
normal leads the quasi-particle pairs are properly separated and their
entanglement can be quantified through a comparison of the intra- and
inter- lead noise.
\begin{figure}
  \centerline{\epsfxsize=7.2cm \epsfbox{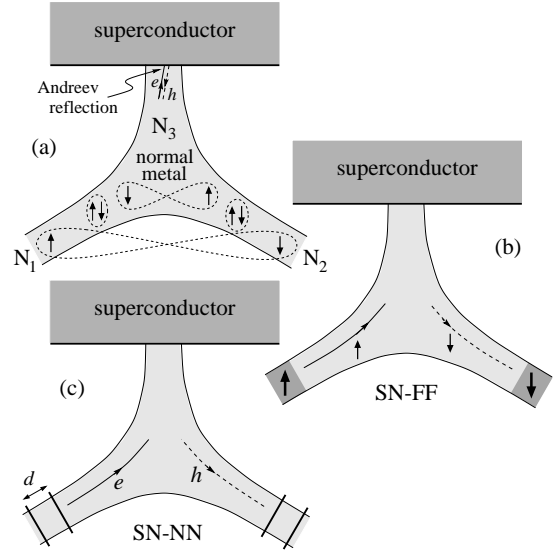}}
  \narrowtext\vspace{4mm} \caption{Normal-metal--superconductor (NS)
    junction with the normal-metal lead arranged in a fork geometry.
    (a) Without filters, entangled pairs of quasi-particles (Cooper
    pairs) injected into the lead N$_3$ propagate into leads N$_1$ or
    N$_2$ either as a whole or one by one.  The ferromagnetic filters
    in setup (b) enforce the separation of the entangled spins, while
    the Fabry-Perot type interferometers in the setup (c) separate
    electron- and hole type quasi-particles.}
\end{figure}
\vskip -.1cm
We start by noting that for a {\it single} channel NS wire the zero
frequency fluctuations of the currents carried by electrons with
different spins are completely correlated, 
\begin{equation}
\langle\langle\, I_{\sigma}I_{-\sigma}\,
\rangle\rangle_{\omega=0}
\equiv \int dt\> \langle\langle I_{\sigma}(t)
I_{-\sigma}(0)\rangle\rangle
=
\langle\langle I^2_{\sigma}\rangle\rangle_{\omega=0},
\label{cc_spin}
\end{equation}
hence $\langle\langle (I_{\sigma}- I_{-\sigma })^2
\rangle\rangle_{\omega=0}=0$ ($\langle\langle...\rangle\rangle$
implies the subtraction of the average currents).  This perfect
correlation in the (subgap) motion of the quasi-particles with
different spins is a consequence of the entanglement of the Cooper
pairs injected into the normal wire.

Next, recall that the current noise cross-correlations in a 
SN-NN fork geometry -- without filters on the normal probes 
-- are positive when the transmission between the superconductor
and the normal leads is low \cite{Torres}. The unusual sign (for 
fermions) of these correlations is due to paired electrons 
penetrating the two normal leads separately, c.f., Fig.\ 1(a). The 
positive correlations are further enhanced when the competing 
channel (with paired electrons entering the leads jointly) is 
suppressed through the addition of appropriate spin- or 
energy selective filters to the normal leads, see  Figs.\ 
1(b) and (c). The wave function of the entangled states 
generated with ideal spin/energy filters then is of the type
\begin{equation}
|\Phi^{\rm ent}_{\varepsilon,\sigma}\rangle =
\alpha |\varepsilon ,\sigma;
-\varepsilon, -\sigma \rangle
+\beta |\mp \varepsilon \pm \sigma;
\pm \varepsilon \mp \sigma \rangle,
\label{wave function}
\end{equation}
where the first (second) argument in $|\phi_1;\phi_2\rangle$
refers to the quasi-particle state in lead 1 (2) evaluated behind
the filters and the upper (lower) signs refer to the setup
projecting the spin (energy); the coefficients $\alpha$ and
$\beta$ can be tuned by external parameters, e.g., a magnetic
field. In such a multi-terminal device, the measurement of the
zero frequency noise cross-correlator $\langle \langle\,I_{\sigma
1} I_{-\sigma 2}\,\rangle\rangle_{\omega=0}$ then serves to detect
electron entanglement, in analogy with the above single channel NS
wire.

%%%%%%%%%%%%%%%%%%%%%%%%%%%%%%%%%%%%%%%%%%%%%%%%%%%%%%%%%%
A step like dependence of the gap parameter at the NS interface is
assumed, subgap transport is specified, while the normal leads are
single-channel and ballistic. Using the scattering formulation of NS
transport\cite{ns_transport}, the current operator per spin in normal
lead $n$ is defined as
\begin{eqnarray}
I_{\sigma n}(t) &=& \frac{ie\hbar}{2m}
\sum_{\alpha,\alpha^\prime}\int_0^\infty \!\! d\varepsilon
d\varepsilon^\prime \{[u^\ast_{\varepsilon\alpha}(x)
\!\stackrel{_\leftrightarrow}{\partial}_x\!
u_{\varepsilon^\prime\!\alpha^\prime}(x)] \,
\gamma_{\varepsilon\alpha}^\dagger
\gamma_{\varepsilon^\prime\!\alpha^\prime}\nonumber \\
&-&[v^\ast_{\varepsilon\alpha}(x)\!
\stackrel{_\leftrightarrow}{\partial}_x \!
v_{\varepsilon^\prime\!\alpha^\prime}(x)] \,
\gamma_{\varepsilon^\prime\!\alpha^\prime}
\gamma_{\varepsilon\alpha}^\dagger\}
\exp[i(\varepsilon-\varepsilon^{\prime})t], \label{def_current}
\end{eqnarray}
where the operators $\gamma_{\varepsilon\alpha}$ describe electron and
hole Bogoliubov quasiparticles (with positive energies $\varepsilon$)
on the normal side with $\alpha = \{p,\sigma,n\}$ a multi-index
characterizing the `charge' $p\,(=e,h)$, spin $\sigma\,(=\pm 1/2)$,
and incidence (lead $n$); $f\stackrel{_\leftrightarrow}{\partial}_x g
\equiv f \partial_x g - g \partial_x f$.  The associated wave
functions $(u_{\varepsilon\alpha}(x)$ and $v_{\varepsilon\alpha}(x))$
are expressed in terms of the scattering matrix $s_{\alpha,
\alpha^\prime}$; e.g., for an electron with spin $\sigma$ incident
from lead $n$ and observed in lead $m$ at $x_m$
\begin{eqnarray}
   u_{\varepsilon e\sigma n}(x_m)
   &\simeq& \Bigl[\delta_{nm}e^{ik_+x_n}
   +s_{e\sigma n,e\sigma m} e^{-ik_+x_m}\Bigr]/\sqrt{h v_+},
   \nonumber \\
   v_{\varepsilon e\sigma n}(x_m)
   &\simeq& \Bigl[s_{e\sigma n,h -\sigma m}e^{ik_-x_m}\Bigr]
   /\sqrt{h v_-}, \label{v_BdG}
\end{eqnarray}
with wave numbers $k_\pm=\sqrt{2m(\mu_{\rm\scriptscriptstyle S} \pm
  \varepsilon)}$ and the quasi-particle velocities $v_\pm=\hbar
k_\pm/m$ ($\mu_{\rm\scriptscriptstyle S}$ is the chemical potential in
the superconductor). The difference between the two wave numbers
$k_\pm$ will be neglected ($\mu_{\rm\scriptscriptstyle S}\gg\Delta$).

%%%%%%%%%%%%%%%%%%%%%%%%%%%%%%%%%%%%%%%%%%%%%%%

Let us now turn to the fork geometry of Fig.\ 1: the current leads
N$_1$ and N$_2$ are connected to N$_3$, which itself is terminated
with the NS interface. We discuss two schemes: a) two ferromagnetic
metal contacts (with magnetizations in opposite directions) in leads
$N_1$ and $N_2$ block the propagation of the opposite spin,
see Fig.\ 1(b), b) two energy filters in $N_1$ and $N_2$ (coherent
quantum dots) select the kinetic energies of electrons and holes
symmetrically above and below the superconducting chemical potential
(Fig.\ 1(c)). In both proposals, the penetration of a Cooper pair
into a given lead is prohibited, while allowing the split pair to
pass the filters. E.g., one electron propagates through
$N_1$ with spin `up', while simultaneously the other electron
(with opposite kinetic energy) propagates through $N_2$ with spin
`down'.

The scattering matrix $s_{\alpha,\alpha^\prime}$ has to account
for all multiple scattering processes: the Andreev reflection at
the N$_3$S interface can be specified in terms of the transmission
and reflection amplitudes $t_{13}$, $t_{23}$, and $r_{ii}$
($i=1,2$) describing the normal-metal part of the device
\cite{Fauchere}. The latter include the scattering by the `beam
splitter' N$_3 \longleftrightarrow$ N$_1$, N$_2$, and account for
the presence of spin or energy filters in leads N$_1$ and N$_2$.
The corresponding transmission and reflection amplitudes are found
iteratively following the scheme sketched in Fig.\ 2 and
accounting for all interference processes in the device.
\begin{figure}
  \centerline{\epsfxsize=5.0cm \epsfbox{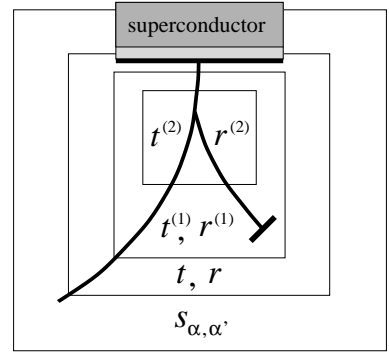}}
  \narrowtext\vspace{4mm} \caption{The scattering matrix
    $s_{\alpha,\alpha^\prime}$ determining the noise correlator 
    (\ref{noise_12}) incorporates
    all the internal scattering features of the fork, the beam
    splitter ($\rightarrow t^{\scriptscriptstyle (2)},~
    r^{\scriptscriptstyle (2)}$), the stubs ($\rightarrow
    t^{\scriptscriptstyle (1)},~r^{\scriptscriptstyle (1)}$), and
    the normal scattering at the NS interface ($\rightarrow t,~r$).}
\end{figure}
{\it Beam Splitter:} Time reversal invariance is assumed
for simplicity.  It is then possible to express the transmission
probability, say, between leads N$_1$ and N$_2$ in terms of the
other two transmissions:
\begin{equation}
   T_{12}^{\scriptscriptstyle (2)}
   = T_{13}^{\scriptscriptstyle (2)}
     T_{23}^{\scriptscriptstyle (2)}
     [2-T_\Sigma^{\scriptscriptstyle (2)}
     \pm 2(1-T_\Sigma^{\scriptscriptstyle (2)})^{1/2}]/
     {T_\Sigma^{\scriptscriptstyle (2)}}^2,
     \label{step_3a}
\end{equation}
where $T_\Sigma^{\scriptscriptstyle (2)} \equiv
T_{13}^{\scriptscriptstyle (2)} + T_{23}^{\scriptscriptstyle
(2)}$, and $T_{ij}^{\scriptscriptstyle(2)}=
|t_{ij}^{\scriptscriptstyle(2)}|^2$.  For a fully symmetric beam
splitter $T_{ij}^{\scriptscriptstyle (2)} = 4/9$. 
Ref.\ \cite{Gefen} focused on a setup which is symmetric between
$1$and $2$. The lower sign in
(\ref{step_3a}) allows to consider also the case where all
transmissions are small, thus a more complete
parametrization is obtained.
The reflection
amplitudes of the splitter are 
\begin{equation}
   r_{ii}^{\scriptscriptstyle (2)}
   = {t_{jk}^{\scriptscriptstyle (2)}}^\ast
   t_{ij}^{\scriptscriptstyle (2)}
   t_{ik}^{\scriptscriptstyle (2)}
   [{T_{jk}^{\scriptscriptstyle (2)}}^{-1}
   -{T_{ij}^{\scriptscriptstyle (2)}}^{-1} -
   {T_{ik}^{\scriptscriptstyle (2)}}^{-1}]/2.
   \label{step_3b}
\end{equation}
The phases of the reflection and transmission amplitudes, in addition
to describing the intrinsic properties of the links between the three
wires, account for the choice of the origin in each lead.
Practical choices for these origins are: a) the
position of the NS interface in lead N$_3$ and b) that of the
filters in N$_1$ and N$_2$.
Typical phases accumulated during free propagation are
$\exp({\pm ik_\pm b})$ with $b$ a typical length in the beam
splitter.

{\it Stubs:} When blocking the propagation through N$_2$ with an
ideal filter, the transmission and reflection amplitudes
$t_{13}^{\scriptscriptstyle (1)}$, $r_{33}^{\scriptscriptstyle
(1)}$, and $r_{11}^{\scriptscriptstyle (1)}$ follow from the
transmission and reflection amplitudes $t_{12}^{\scriptscriptstyle
(2)}$, $t_{13}^{\scriptscriptstyle (2)}$,
$t_{23}^{\scriptscriptstyle (2)}$, $r_{11}^{\scriptscriptstyle
(2)}$, and $r_{22}^{\scriptscriptstyle (2)}$ of the three bare
leads via
\begin{mathletters}
\begin{eqnarray}
   t_{13}^{\scriptscriptstyle (1)}
   &=&t_{13}^{\scriptscriptstyle(2)}
-(t_{12}^{\scriptscriptstyle (2)}
t_{23}^{\scriptscriptstyle (2)}    -t_{13}^{\scriptscriptstyle
(2)} r_{22}^{\scriptscriptstyle (2)})/   
(1+r_{22}^{\scriptscriptstyle (2)}),    \label{step_2t} \\    
   r_{ii}^{\scriptscriptstyle (1)}
   &=&r_{ii}^{\scriptscriptstyle (2)}
   -t_{2i}^{\scriptscriptstyle (2)2}/
   (1+r_{22}^{\scriptscriptstyle (2)})~~~~~,~~(i=1,3)
   \label{step_2r}
\end{eqnarray}
\end{mathletters}
When blocking occurs at $N_1$, the amplitudes are obtained by
exchanging the lead indices in (\ref{step_2t}) and
(\ref{step_2r}).

{\it NS boundary:} The NS boundary is split
into two parts describing normal and Andreev scattering
(Fig.\ 2). In order to include the normal scattering component 
the two scatterers $\{t_{13}^{\scriptscriptstyle (1)},
r_{33}^{\scriptscriptstyle (1)}, r_{11}^{\scriptscriptstyle
(1)}\}$ and $\{t_{\rm\scriptscriptstyle NS}, r_{\rm
\scriptscriptstyle NS}, r_{\rm\scriptscriptstyle NS}^\prime\}$
are combined to obtain the next level amplitudes:
\begin{mathletters}
\begin{eqnarray}
   t_{13}
   &=& t_{13}^{\scriptscriptstyle (1)}
   t_{\rm\scriptscriptstyle NS}/
(1-r_{33}^{\scriptscriptstyle
(1)}    r_{\rm\scriptscriptstyle NS}^\prime),
   \label{step_1t} \\
   r_{33}
   &=& r_{\rm\scriptscriptstyle NS} +
   t_{\rm\scriptscriptstyle NS}^2
   r_{33}^{\scriptscriptstyle
(1)}/
(1-r_{33}^{\scriptscriptstyle (1)}   
r_{\rm\scriptscriptstyle NS}^\prime),  
\label{step_1tt} \\
   r_{11}
   &=&{r_{11}^{\scriptscriptstyle (1)}}
   +t_{13}^{\scriptscriptstyle (1)2}
   r_{\rm\scriptscriptstyle NS}^\prime/
   (1-r_{33}^{\scriptscriptstyle (1)}
   r_{\rm\scriptscriptstyle NS}^\prime),
   \label{step_1r}
\end{eqnarray}
\end{mathletters}
where the primed reflection amplitude $r'_{\rm\scriptscriptstyle
NS}$ is associated with the particle incident from the
superconductor. Finally, we account for perfect Andreev reflection
at the NS interface; for an electron incident from N$_1$ the
scattering amplitude reads
\begin{equation}
   |s_{e\sigma 1,h -\sigma 2}|^2
   = \frac{|t_{13+}|^2|t_{23-}|^2}{1+|r_{11+}|^2 |r_{22-}|^2
   + 2{\rm Re} (r_{33+}r_{33-}^\ast)},
   \label{step_0}
\end{equation}
where the indices $\pm$ indicate that the energy dependent
scattering amplitudes have to be evaluated at the
positive(negative) value of the quasi particle energy
$\varepsilon$(-$\varepsilon$).  For each energy $\varepsilon$ only
one of the two leads, N$_1$ or N$_2$, is open, resulting in a
two-terminal device, thus the relations $|r_{ii}|^2=1-|t_{i3}|^2 =
|r_{33}|^2$ hold. The main feature contained in Eq.\
(\ref{step_0}) are the Andreev type resonances building up within
the normal-metal leads. These resonances are determined through
the sign changes in ${\rm Re} (r_{33+}r_{33-}^\ast)$ and their
distance $\sim \hbar v_{\rm\scriptscriptstyle F}/L$ is determined
via the Fermi velocity $v_{\rm\scriptscriptstyle F}$ and the
characteristic size $L$ of the region. In addition, zeros appear
in the spectral density which are a consequence of a vanishing
transmission for electrons or holes in this three lead geometry.

The above scheme fully specifies the scattering matrix $s_{\alpha,
\alpha^\prime}$ for the case with ideal filters. For non-ideal
filters the stub should be replaced with a proper description of
the lead N$_2$ including its non-ideal filter; in addition, the
normal scatterer described through the amplitudes
$\{t_{13},r_{33},r_{11}\}$ above has to be combined with an
additional scatterer describing the non-ideal filter in the lead
N$_1$. E.g., an energy filter requires inclusion of a Fabry-Perot
interferometer characterized through the scattering amplitudes
$t_1,t_2,r^\prime_1,r_2$ and the separation $d$ of the double
barrier system and producing a transmission $t_{\rm
\scriptscriptstyle FP} = t_1 t_2 \exp(ikd)/[1-r_1 r^\prime_2
\exp(2ikd)]$.  The resonance spacing should be larger than the
applied bias for a proper device operation as a filter. The
initial resonance lines produced by the quantum dot will then be
decorated by Andreev-type resonances and zeros originating from
the NS-fork structure.

%%%%%%%%%%%%%%%%%%%%%%%%%%%%%%%%%%%%%%%%%%%%%%%

The above entangler is essentially a two terminal device where
electrons with, say, a given spin from lead $1$ are converted into
holes with an opposite spin in lead $2$. The current correlations
between $1$ and $2$ are positive and can be obtained using the
definition of the noise in combination with (\ref{def_current}),
at $T=0$,
\begin{equation}
   \langle\langle I_{\sigma 1} I_{-\sigma 2} \rangle\rangle
   =\frac{e^2}{h}\sum_{\alpha,\alpha^\prime}\int_0^{eV_1}
   d\varepsilon \, |s_{\alpha,\alpha^\prime}|^2
   (1-|s_{\alpha,\alpha^\prime}|^2),
   \label{noise_12}
\end{equation}
where a voltage $eV_1$ is applied between the lead N$_1$ and the
superconductor while keeping the lead N$_2$ unbiased. For the case
of ferromagnetic filters, the chemical potential which enters the
(sharp) electron and hole distribution functions depends also on
the spin index. The multi-indices $\alpha$ and $\alpha^\prime$ to
be summed over in (\ref{noise_12}) depend on the type of filters
in the normal leads N$_1$ and N$_2$: For ferromagnetic filters
(SN-FF) with the spin in F$_{1(2)}$ pointing up (down) $\alpha =
\{e(h) \uparrow 1\}$ and $\alpha^\prime = \{h(e) \downarrow 2\}$
(the propagation of other states is blocked by the filters). On
the other hand, for the setup selecting a definite quasi particle
energy via Fabry-Perot type filters we have to sum over spins with
$\alpha = \{e \uparrow\!(\downarrow)\, 1\}$ and $\alpha^\prime =
\{h \downarrow\!(\uparrow)\, 2\}$ (we assume filters selecting
quasi particles and quasi holes in leads N$_1$ and N$_2$,
respectively). Applying the same voltage to the lead N$_2$ as well
does not change the answer in the normal fork (SN-NN) but renders
the result for the ferromagnetic filters (SN-FF) twice larger.
The
current fluctuations (\ref{noise_12}) are straightforwardly
converted into ``counting'' correlations as known from quantum
optics: with $e n_{\sigma n}(t) = \int_0^t dt^\prime I_{\sigma
n}(t^\prime)$ we find: 
\begin{equation}
\langle\langle n_{\sigma 1}(t)\, n_{-\sigma
2}(t)\rangle\rangle|_{t\to\infty}\approx (t/e^2)\langle\langle
I_{\sigma 1}\,I_{-\sigma 2}\rangle \rangle_{\omega=0}~.
\label{irreducible_number}
\end{equation} 
and hence
$\langle\langle(n_{\sigma 1}-n_{-\sigma 2})^2\rangle\rangle
/\langle\langle n_{\sigma 1}^2\rangle\rangle \approx 0$.
The result (\ref{noise_12}) together with the fact
that the two currents $I_{\sigma 1}$ and $I_{-\sigma 2}$ are
necessarily correlated constitutes the main justification for the
proposed entanglement device: Eq.\ (\ref{noise_12}) corresponds
precisely to the current noise in lead $1$. Ideally, these
correlation measurements should be performed using fast
electronics in order to generate time resolved voltage pulses for
electron injection/detection.

This electronic entanglement apparatus can now be completed with a
detection apparatus in order to test non-local correlations (Bell
inequalities). For an entangler based on ferro/energy filters, the
detection apparatus involves filters of opposite type
(energy/ferro). Concentrating on energy filters, a positive energy
particle emerging in lead N$_1$ can have either spin orientation,
which can be ``measured'' by connecting this lead to, say, a
magnetic contact with known spin orientation. In the opposite arm
one should have a similar contact but with a magnetization axis
rotated by $\pi/2$ in order to achieve the analog of the spin
correlation experiments of Ref.\ \cite{ent_photons}.
A detailed discussion of these will be provided later
\cite{Chtchelkatchev}.

The proximity induced entanglement of quasi particles in NS-fork
type devices was implicit in Refs.\ \cite{Tsoi} and 
Ref.\ \cite{Feinberg}. Consider the above SN-NN
setup with the lead N$_1$ biased with respect to the
superconductor, while keeping the lead N$_2$ at the
superconductor chemical potential. A 
finite current $I_2(V_1)=(2e/h)\int^{eV_1}_{0}
d\varepsilon \left|s_{e\uparrow 1,h\downarrow 2}\right|^2$ will
flow through lead N$_2$ in response to the bias $eV_1$ across
lead N$_1$.  While both experiments in Ref.\ \cite{Tsoi} use a
magnetic field to separate electron- and hole type quasi
particles, the more recent suggestion in Ref.\ \cite{Feinberg}
proposes two ferromagnetic needles, a setup similar to our SN-FF
device.  This Andreev drag effect is quite robust and decreases
only gradually with decreasing quality of the filters. The
condition for this drag effect to persist reads: $\int^{eV_1}_{0}
d\varepsilon [ \left|s_{e\uparrow 1,h\downarrow 2}\right|^2
-\left|s_{e\uparrow 1,e\uparrow 2}\right|^2] >0$, implying that
the normal current injected from lead N$_1$ to lead N$_2$
remains smaller than the `drag current' due to Andreev reflected
holes. Note that the current $I_2$ will vanish when replacing
the $s$-wave superconductor with a $p$-wave material or a normal
conductor.

Summarizing, we have proposed an electronic entanglement device
based on the proximity effect and have shown how to probe the
resulting non-local electronic correlations in an emphatic way
through a measurement of the current cross-correlator. Using a
special fork geometry with, say, Fabry-Perot filters one arrives
at a natural source of spin-entangled electron pairs, a device
with potential applications in quantum computing architectures
based on spintronics\cite{LossdiVincenzo}. This device presents
the advantage -- as compared to its ferromagnetic cousin -- that
it can be realized with nowadays splitters \cite{Henny_Oliver} and
quantum dot technology, e.g., using semiconductor-superconductor
heterostructures \cite{Takayanagi}, and does not require
interdot coupling.
Moreover, this SN-NN device appears to be more promising
regarding potential applications for quantum information
processing: the insertion of Fabry-Perot filters destroys only
the orbital entanglement of the electrons, while the (most
valuable) spin entanglement persists, contrary to the situation
in the SN-FF device where the filters {\it project the spin}, but
where the entanglement of {\it energy degrees of freedom}
persists nevertheless.

We thank M.\ Feigelman and M.\ Reznikov for useful discussions and the
Swiss NSF for financial support. GBL acknowledges support from a NWO
grant for collaboration with Russia.

\end{multicols}
\end{document}